\documentclass[aps,reprint,letterpaper,amsmath,amssymb, showpacs, superscriptaddress, longbibliography]{revtex4-1}
\usepackage{graphicx}
\usepackage[absolute]{textpos}

\begin{document}
\begin{textblock*}{8.5in}(0.1in,0.25in)
\begin{center}
JETP Letters \textbf{101}, 513 (2015)
\end{center}
\end{textblock*}
\begin{textblock*}{2.5in}(5.06in,4.21in)
{\small \doi{10.1134/S0021364015080111}}
\end{textblock*}

\title{Dynamics of spatial coherence and momentum distribution
of polaritons in a semiconductor microcavity under conditions
of Bose-Einstein condensation}

\author{D. A. Mylnikov}
\affiliation{Lebedev Physical Institute, Russian Academy of Sciences, Moscow, 119991 Russia}
\affiliation{Moscow Institute of Physics and Technology (State University), Dolgoprudnyi, Moscow region, 141700 Russia}
\author{V. V. Belykh}
\email[]{belykh@lebedev.ru}
\affiliation{Lebedev Physical Institute, Russian Academy of Sciences, Moscow, 119991 Russia}
\author{N. N. Sibeldin}
\affiliation{Lebedev Physical Institute, Russian Academy of Sciences, Moscow, 119991 Russia}
\author{V.~D.~Kulakovskii}
\affiliation{Institute of Solid State Physics, Russian Academy of Sciences, Chernogolovka, Moscow region, 142432 Russia}
\author{C. Schneider}
\affiliation{Technische Physik, Physikalisches Institut and Wilhelm Conrad R\"{o}ntgen Research Center for Complex Material Systems, Universit\"{a}t W\"{u}rzburg, D-97074 W\"{u}rzburg, Germany}
\author{S. H\"{o}fling}
\affiliation{Technische Physik, Physikalisches Institut and Wilhelm Conrad R\"{o}ntgen Research Center for Complex Material Systems, Universit\"{a}t W\"{u}rzburg, D-97074 W\"{u}rzburg, Germany}
\author{M. Kamp}
\affiliation{Technische Physik, Physikalisches Institut and Wilhelm Conrad R\"{o}ntgen Research Center for Complex Material Systems, Universit\"{a}t W\"{u}rzburg, D-97074 W\"{u}rzburg, Germany}
\author{A. Forchel}
\affiliation{Technische Physik, Physikalisches Institut and Wilhelm Conrad R\"{o}ntgen Research Center for Complex Material Systems, Universit\"{a}t W\"{u}rzburg, D-97074 W\"{u}rzburg, Germany}

\date{10 March 2015}

\begin{abstract}
The dynamics of spatial coherence and momentum distribution of polaritons in the regime of Bose-Einstein condensation are investigated in a GaAs microcavity with embedded quantum wells under nonresonant excitation with picosecond laser pulses. It is shown that the onset of the condensate first order sparial coherence is accompanied by narrowing of the polariton momentum distribution. At the same time, at sufficiently high excitation densities, there is significant qualitative discrepancy between the dynamic behavior of the width of the polariton momentum distribution determined from direct measurements and that calculated from the coherence spatial distribution. This discrepancy is observed at the fast initial stage of the polariton system kinetics and, apparently, results from the strong spatial nonuniformity of the phase of the condensate wave function, which equilibrates on a much longer time scale.
\end{abstract}


\maketitle

Polaritons in semiconductor microcavities with embedded quantum wells (QWs), are quasiparticles corresponding to mixed exciton-photon states in these heterostructures, represent a unique system for studying Bose-Einstein condensation (BEC). The effective mass of microcavity polaritons near the bottom of the lower dispersion branch is $\sim10^{-4}$ of the free-electron mass, which makes possible polariton BEC at high temperatures, even up to the room temperature \cite{Imamoglu}. After the demonstration of BEC in CdTe microcavities in 2006 \cite{Kasprzak06}, a number of interesting effects related to this phenomenon were investigated, including quantum vortices \cite{Lagoudakis08}, the superfluidity of the polariton condensate \cite{Amo09}, the Josephson effect \cite{Lagoudakis10}, the spin-Meissner effect \cite{Larionov10}, polariton solitons \cite{Sich12}, and the optical spin Hall effect \cite{Kammann12}. Very recently, an electrically pumped polariton laser working at room temperature was demonstrated \cite{Bhattacharya14}.

One of the main characteristics of a Bose condensate is spatial coherence, i.e., the constancy of the phase of the condensate wavefunction over distances exceeding the thermal de Broglie wavelength. In this context, the dynamics of the onset of coherence in the process of the condensate formation is an interesting issue, which was investigated in \cite{Nardin09, Ohadi12, Belykh13}. In particular, in Ref.~\cite{Belykh13} it was found that, in the process of formation of the polariton condensate, coherence expands with a constant velocity of $\sim 10^8$\,cm/s.

In the present work, for the first time, the dynamics of the spatial distribution of coherence and the corresponding
dynamics of the polariton momentum distribution are investigated simultaneously. The relationship between these distributions is discussed. The structure under study is a GaAs microcavity with the top and bottom Bragg mirrors consisting of 32 and 36
pairs of AlAs/Al$_{0.13}$Ga$_{0.87}$As layers, respectively. 12 GaAs/AlAs 7-nm-wide QWs are embedded in the microcavity. The Rabi splitting and the $Q$ factor of the structure are 4.5 meV and about 7000, respectively. The detuning between the photon and exciton modes is about -6 meV. In all experiments, the sample temperature is 10 K. The sample is excited by a Ti:sapphire laser emitting a periodic train of 2.5-ps-long pulses at a repetition rate of 76 MHz. The excitation laser beam is focused in a spot of 20-30 $\mu$m in diameter. The excitation photon energy is 10-20 meV higher than the energy of the bottom of the exciton mode.

The degree of spatial coherence is measured using the Young double-slit experiment scheme \cite{Belykh13, Deng07}. Using two converging lenses ($F_1\approx 1$ and $F_2\approx 10$\,cm) with coinciding focal planes, an enlarged image of the sample with a magnification of about 13 is obtained. A metal-coated plate with a pair of 5-$\mu$m-wide transparent slits is placed in the plane of this image. The slit spacing can be varied from 20 to 160 $\mu$m. The chosen slit pair selects two regions in the image of the spot, and the pattern formed by the interference of radiation emitted from these two regions is imaged by a third lens onto the entrance slit of a spectrometer set to the zeroth diffraction order and coupled to a streak-camera.

To measure the distribution of polaritons in momentum space ($k$ space), the plate with slits is removed from the above setup. Then, owing to the presence of the third lens, radiation emitted by the microcavity at a certain angle $\theta$ to the normal of the
sample surface and, thus, corresponding to polaritons with a wave vector $k=(\omega/c) \sin(\theta)$ is imaged onto a
certain point at the slit of the spectrometer and, thus, of the streak camera. The time resolution of the system in our measurements is 10~ps.

Another quantity determined in our experiments is the total number of polaritons $N$ occupying the states near the bottom of the lower polariton branch (LPB). This is attained by measuring the power of radiation emitted by the sample. The procedure of determining $N$ is described in detail in  \cite{Belykh13}.

\begin{figure*}
\begin{center}
\includegraphics[width=2\columnwidth]{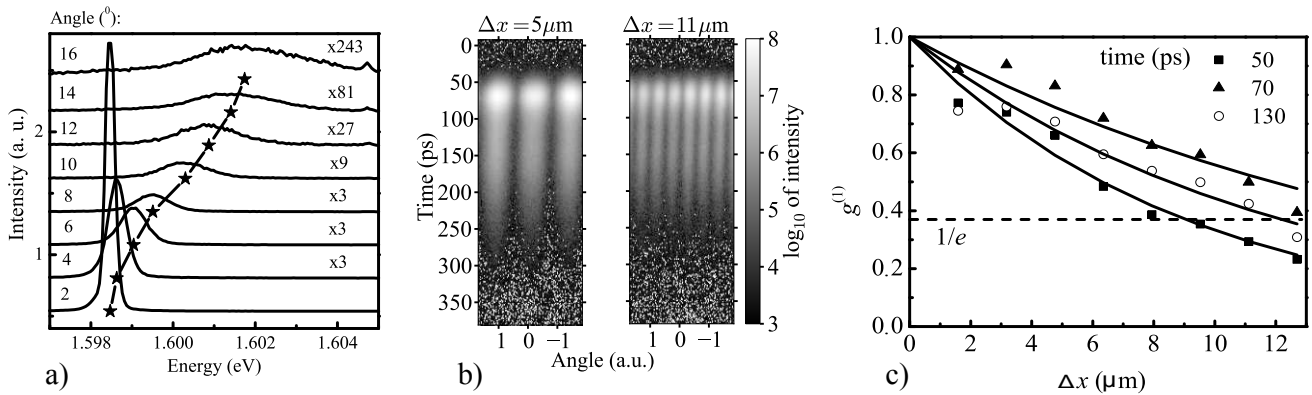}
\caption{(a) Time-integrated PL spectra of the sample recorded at different angles with respect to its normal. Asterisks show the positions of the spectral lines. The pump power is $P = 1.2P_0$. (b) The dynamics of the interference pattern for different distances $\Delta x$ between the selected regions of the emitting spot. The pump power is $P = 3.4P_0$. (c) (Symbols) Coherence function versus the distance $\Delta x$ at different times. Solid lines show the fits to these dependencies by exponential functions. The pump power is $P = 3.4P0$.}
\end{center}
\end{figure*}
As the pump power is increased above $P_0\approx 750$~W/cm$^2$ a threshold-like increase in the intensity
of photoluminescence (PL) emitted at small angles with respect to the normal takes place, giving evidence
of the onset of the macroscopic occupancy of the states near the LPB bottom. This is illustrated by Fig. 1a, which
shows the PL spectra recorded at different angles at a pump density $P=1.2P_0$. The positions of the spectral
lines are shown by asterisks. The angular dependence of these positions describes a part of the dispersion
curve corresponding to the LPB states in the strong-coupling regime.

Above the threshold pump density, measurements in the Young scheme reveal an interference pattern
formed by radiation emitted from the regions of the spot selected by the two slits. This gives evidence of the
onset of spatial coherence and the BEC of polaritons. The dynamics of the interference pattern for the separations between the emitting regions of $\Delta x=5$ and 11~$\mu$m are shown in Fig.1b. In these experiments, the sample image is adjusted with respect to the slits in such a way that the intensities of light collected from each of the slits are the same. Then, the degree of
spatial coherence $g^{(1)}$ is equal to the interference pattern visibility, which is calculated by performing a Fourier transform. The dependences of the coherence function on the distance $\Delta x$ for different times after the excitation pulse are shown in Fig. 1c. The coherence length $r_\text{c}$ is determined as the distance at which $g^{(1)}(r_\text{c})=1/e$. To calculate $r_\text{c}$ the dependence of $g^{(1)}(\Delta x)$ is approximated by exponential functions, also shown in Fig. 1c.

\begin{figure*}
  \includegraphics[width=1\linewidth]{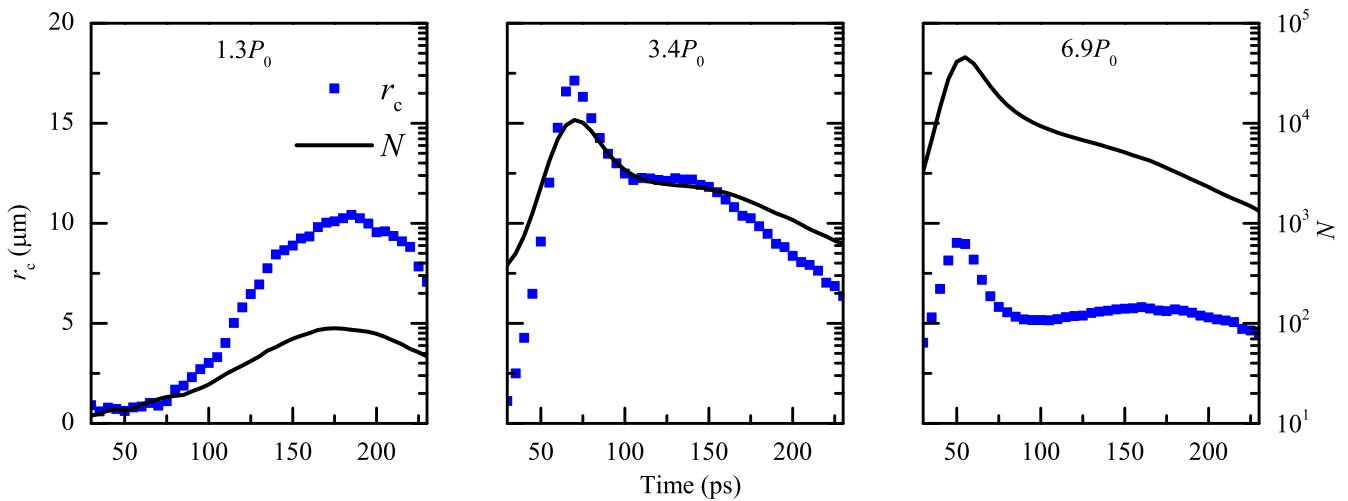}\\
  \caption{Kinetics of the coherence length (symbols, left scale) and the total number of polaritons near the bottom of the LPB (lines, right scale) for different pump powers.}
\label{fig2_rc_N}
\end{figure*}
Figure 2 shows the kinetic dependences of the coherence length $r_\text{c}$ and the total number $N$ of polaritons near the bottom of the LPB. For excitation densities slightly exceeding the BEC threshold density $P_0$, the slow rise and decay of $N$ with a characteristic time of 50 ps is accompanied by qualitatively similar variation in $r_\text{c}$. For densities considerably higher than $P_0$, the kinetics of $N$ clearly features two regions: the initial stage with a total duration of about 50~ps and the second stage of slower decay. At the first (initial) stage, the kinetics of $r_\text{c}$ follows that of $N$: an increase in $N$ is accompanied by an increase in $r_\text{c}$, and a drop in $N$ is accompanied by a decrease in $r_\text{c}$. A characteristic
maximum in the kinetics of $r_\text{c}$ is achieved simultaneously with the maximum in the number of polaritons. At the second stage, the dynamics of $r_\text{c}$ changes abruptly. There appears a plateau, i.e., an interval of time during which $r_\text{c}$ shows almost no decrease. This interval is as long as 100~ps for $P = 6.9P_0$. At longer times, $r_\text{c}$ decays together with $N$, which decreases with a characteristics time of 60 ps. As the pump power is increased, the maximum value of $r_\text{c}$ attained in the process of the condensate formation first increases, has a maximum
for $P = 2.3P_0$, and decreases at higher powers. Note, that the maximum value of $r_\text{c}$ is almost equal to the size of the PL spot. This means that the condensate occupies the entire excitation region. The decrease in $r_\text{c}$ at high excitation powers can be related both to an increase in the energy of particle-particle interactions in the condensate \cite{Belykh13} and to the stronger
overheating of the exciton reservoir \cite{Belykh14}.

\begin{figure*}
  \includegraphics[width=1\linewidth]{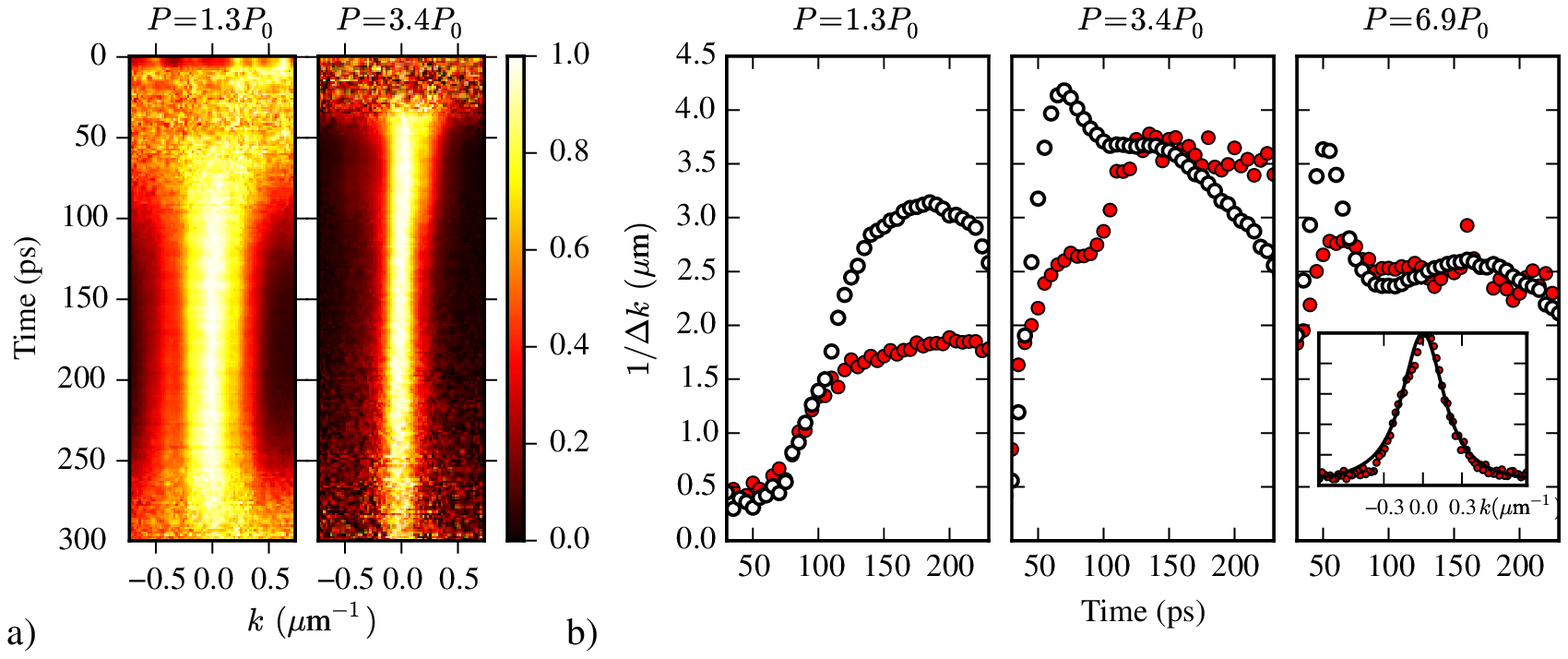}\\
  \caption{(a) Kinetics of the polariton distribution in $k$ space for pump powers $P = 1.3P_0$ and $3.4P_0$. The distribution is normalized to the intensity maximum at each moment of time. (b) The kinetics of the inverse width of the polariton distribution in $k$ space measured experimentally (closed symbols) and calculated from the coherence function according to Eq.~\eqref{kg} (open symbols). The inset shows the measured and calculated function $I(k)$ (symbols and solid line, respectively) for P = 6.9P0 and t = 150 ps.}
  \label{fig3_k_dk}
\end{figure*}

The expansion of coherence and the formation of a condensate are accompanied by narrowing of the polariton momentum distribution. This is illustrated by Fig. 3a, which shows the dynamics of the polariton momentum distribution for excitation densities of
$1.3P_0$ and $3.4P_0$. The kinetics of the inverse width (FWHM) $1/\Delta k$ of the momentum distribution for three different excitation densities are shown in Fig. 3b by closed symbols. In an ideal Bose gas at equilibrium, the polariton momentum distribution function $f(k)$ and the spatial distribution of coherence $g^{(1)}(\Delta x)$ are connected by the Fourier transform \cite{Deng07}; in this way a relation between $r_\text{c}$ and $\Delta k$ can be obtained. For a classical two-dimensional Bose gas below the BEC threshold, $f(k)\sim \exp(-k^2/2mT)$, $g^{(1)}(\Delta x)=\exp(-mT\Delta x^2/2)$ and correspondingly $r_c
\Delta k =3.3$; above the BEC threshold $f(k) \sim 1/(\exp(k^2/2mT-\mu/T)-1)\approx 1/(k^2/2mT-\mu/T)$, $g^{(1)}(\Delta
x) \sim \exp(-\Delta x\sqrt{2m|\mu|})$, and correspondingly $r_c\Delta k \approx 2$. However, the product of $r_c$ and $\Delta k$, measured experimentally are significantly larger than 3.3 at the time corresponding to the highest polariton density and decreases to 2 (for $P = 6.9P_0$) at the slow stage of the condensate kinetics. Apparently, this is caused by the strongly nonequilibrium character and the inhomogeneity of the system under consideration, as well as the nonideality of the polariton gas.

In the general case, the relation between the polariton momentum distribution and the spatial distribution of coherence depends on the size of the system and the spatial distribution of the phase $\varphi(x)$ of the condensate. The dependence of the far-field radiation intensity on the wave vector $k$, which (disregarding the small variation of the photon Hopfield coefficient near the bottom of the LPB) represents the momentum distribution of polaritons, is determined by the interference of electric fields corresponding to different points of the condensate and is given by the formula
\begin{multline}
I(k)=\int_{-\infty}^{\infty} \int_{-\infty}^{\infty} E(x_1) E(x_2) g^{(1)}(|x_2-x_1|) \times
\\ \times \exp\bigl[ik(x_2-x_1) + i\varphi(x_2)-i\varphi(x_1)\bigr]dx_1
dx_2. \label{kg}
\end{multline}
This expression is obtained by the summation of the electric field amplitudes $E(x)$ along the PL spot and time averaging. Note that, in general, one needs to consider the integration with respect to two dimensions, $x$ and $y$. However, it can be shown that, if the function $E(x, y)$ allows the separation of variables (e.g., when it is a Gaussian), this integral is reduced to
Eq.~\eqref{kg}, where $g^{(1)}$ is defined as the degree of coherence measured in a double-slit experiment. Equation \eqref{kg} can also be used to calculate the intensity distribution in the experiment with two infinitely narrow slits separated by $\Delta x$ by substituting a field distribution of the form $E(x)=E(\Delta x/2)[\delta(x-\Delta x/2)+\delta(x+\Delta x/2)]$.

We calculated the distribution $I(k)$ at different times under the following assumptions: the spatial coherence function can be written as $g^{(1)}(x) = \exp(-x/r_\text{c})$, where $r_\text{c}$ is determined from the experiment; the phase distribution is uniform, i.e., $\varphi(x) = const$; and the spatial distribution of the electric-field amplitude of the condensate is described by a Gaussian $\exp(-x^2/ L^2)$. The size $L$ of the sample area occupied by the condensate is determined by measuring
the size of the magnified image of the condensate. For $P = 1.3P_0$, $3.4P_0$, and $6.9P_0$, this size is $L =$ 12, 14,
and 20 $\mu$m, respectively. The measured and calculated distributions $I(k)$ at time $t = 150$~ps for $P = 6.9P_0$ are
shown in the inset in Fig.~3b. The kinetic dependencies of the inverse width $1/\Delta k$ characterizing the calculated distributions $I(k)$ are shown in Fig. 3b by open symbols. As one would expect, they reproduce to a large extent the kinetics of $r_c$ (Fig.~2). At the same time, there are significant differences between the measured and calculated kinetic dependencies of $1/\Delta k$ (Fig.~3b).

At a low excitation density of $1.3P_0$, the calculated and observed kinetics of $1/\Delta k$ are similar, but the calculated values exceed the measured ones by almost a factor of 2. At higher excitation densities, there are significant differences between the behaviors of the observed and calculated kinetics. For $P = 3.4P_0$, the measured dependence of $1/\Delta k$ exhibits at the first (fast) stage of the kinetics a short plateau followed by a rapid rise before entering a longer plateau at the second (slow) stage. At the same time, the calculated dependences of $1/\Delta k$ exhibit a sharp maximum at the fast stage of the kinetics and then, after an abrupt drop, enter a plateau at the slow stage (see Fig.~3b for $P = 3.4P_0$ and $6.9P_0$). The differences in the behaviors of the decay kinetics of $1/\Delta k$ for $t > 200$~ps are probably caused by the experimental uncertainty owing to the
small PL intensity in this time range.
\begin{figure}
  \includegraphics[width=1\linewidth]{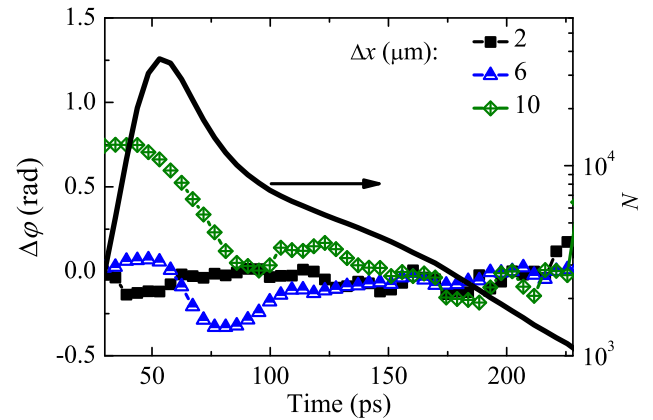}\\
  \caption{Kinetics of the phase difference $\Delta \varphi$ between the areas of the condensate selected by the two slits for different distances $\Delta x$ (left scale, symbols) and the number of polaritons $N$ near the bottom of the LPB (right scale, solid line). The phase difference is measured with respect to its value at time $t = 230$~ps. The excitation power is $P = 6.9P_0$.}
\end{figure}

The discrepancies in the values of $1/\Delta k$ measured directly and calculated on the basis of the experimentally determined dependences $g^{(1)}(x)$ that occur at the fast stage of the kinetics are most probably related to the strong spatial nonuniformity of the phase of the condensate wavefunction in this time range, which is disregarded in the calculations. This nonuniformity causes a phase shift $\Delta \varphi$ between the areas of the condensate selected by the two slits, which, in the measurements of $g^{(1)}(x)$, leads to a shift of the interference pattern without affecting its visibility.
At the same time, phase nonuniformity causes significant broadening of the distribution $I(k)$. This can easily be seen upon substituting, e.g., a linearly varying phase $\varphi(x) \sim x$ in Eq.~\eqref{kg}. This assumption is supported by the data in Fig. 4, which shows the kinetic dependences of the phase difference $\Delta \varphi$ between the regions of the condensate selected by the two slits. The values of $\Delta \varphi$ are determined from the shift of the interference pattern. At the fast stage of the kinetics, $\Delta \varphi$ varies considerably, and these variations are larger for larger distances $\Delta x$ between the corresponding areas of the condensate. These results give evidence of the strong spatial nonuniformity of the phase. Upon the transition to the slow stage of the kinetics, the variation of $\Delta \varphi$ becomes much weaker, and the phase distribution tends to the steady state.

The spatial nonuniformity of the phase may originate from the inhomogeneity of the potential energy of polaritons in the microcavity plane, which may cause the formation of topological defects in the condensate, such as quantum vortices \cite{Lagoudakis08} and dark solitons \cite{Demenev14}. Upon a round trip about the core of such a defect, the phase varies by $2\pi$, and the defect size is determined by the healing length $\xi = \hbar/\sqrt{(2 m \alpha N/S)}$
\cite{Demenev14}, where $m$ is the polariton effective mass, $\alpha$ is the polariton-polariton interaction constant, and $S$ is the area of the condensate. Thus, $\xi$ determines the characteristic length scale upon which the phase varies by $\sim \pi$. Substituting numerical values of $m = 5 \times 10^{-32}$~g, $\alpha = 10^{-12}$~meV cm$^2$, and $S = 10^{-6}$ cm$^2$, we find that, for $P =3.4P_0$, near the peak of the intensity, when $N \sim 10^4$, the healing length is $\xi = 8$ $\mu$m noticeably
smaller than the size of the area occupied by the condensate ($L = 14$ $\mu$m). Thus, the phase nonuniformity resulting from the formation of topological defects may lead to the considerable broadening of the polariton momentum distribution. At the slow stage of the kinetics, $N \sim 2000$ and $\xi = 18$ $\mu$m. In this situation, the contribution of phase nonuniformity to the increase in $\Delta k$ is much smaller.

Note that, at higher excitation densities, the formation of the condensate may take place in several areas of the microcavity simultaneously, which may also contribute to the initial nonuniformity of the phase. Indeed, phases in these areas are initially independent and equalize in the course of the condensate formation.

In conclusion, we have for the first time simultaneously measured the dynamic characteristics of the spatial coherence and polariton momentum distribution under the conditions of polariton Bose-Einstein condensation in a GaAs microcavity with embedded QWs. We have shown that, at sufficiently high densities of nonresonant picosecond laser excitation, there are two stages in the dynamics of the total number of polaritons in the condensate. At the first, fast, stage, the coherence length increases abruptly and then decreases, following the corresponding changes in the number of polaritons occupying states near the bottom of the LPB. At the same time, the polariton momentum distribution exhibits steady narrowing that ceases only at the slow stage in the kinetics of the number of particles in the condensate. The observed difference in the character of the dynamics of spatial coherence and polariton momentum distribution at the fast stage of the condensate kinetics is presumably caused by spatial nonuniformity in the distribution of the phase of the condensate wavefunction, which smoothens much slower.

We are grateful to S.S. Gavrilov, N.A. Gippius, and S.I. Novikov for valuable advice and useful discussions. The work was supported by the Russian Foundation for Basic Research (project nos. 12-02-33091, 13-02-12197, and 14-02-01073) and the Presidium of the Russian Academy of Sciences. The work of V.V.B. was partly supported by the Russian Federation President Scholarship.

\end{document}